# Nano-ARPES investigation of twisted bilayer tungsten disulfide

Giovanna Feraco[1], Oreste De Luca[1†], Przemysław Przybysz[1,2], Homayoun Jafari[1], Oleksandr Zheliuk[1], Ying Wang[1], Philip Schädlich[3], Pavel Dudin[4], José Avila[4], Jianting Ye[1], Thomas Seyller[3], Paweł Dąbrowski[2], Paweł Kowalczyk[2], Jagoda Sławińska[1*], Petra Rudolf[1*], Antonija Grubišić-Čabo[1*].

[1] Zernike Institute for Advanced Materials, University of Groningen, Nijenborgh 4, 9747 AG Groningen, The Netherlands

[2] Faculty of Physics and Applied Informatics, University of Łódź, Pomorska 149/153, 90-236 Łódź, Poland

[3] Institut für Physik, Technische Universität Chemnitz, Reichenhainer Str. 70, 09126 Chemnitz, Germany

[4] Synchrotron SOLEIL, Université Paris-Saclay, Saint Aubin, BP 48, 91192 Gif sur Yvette, France

*Corresponding authors. E-mail: a.grubisic-cabo@rug.nl, p.rudolf@rug.nl, jagoda.slawinska@rug.nl. AGC and PR for ARPES measurements; JS for DFT calculations.

† Current address: Dipartimento di Fisica, Università della Calabria, Via Pietro Bucci Cubo 31C, 87036 Arcavacata di Rende (CS), Italy

## Abstract

The diverse and intriguing phenomena observed in twisted bilayer systems, such as graphene and transition metal dichalcogenides, prompted new questions about the emergent effects that they may host. However, the practical challenge of realizing these structures on a scale large enough for spectroscopic investigation, remains a significant hurdle, resulting in a scarcity of direct measurements of the electronic band structure of twisted transition metal dichalcogenide bilayers. Here we present a systematic nanoscale angle-resolved photoemission spectroscopy investigation of bulk, single layer, and twisted bilayer $WS_2$ with a small twist angle of 4.4°. The experimental results are compared with theoretical calculations based on density functional theory along the high-symmetry directions $\bar{\Gamma}-\bar{K}$ and $\bar{\Gamma}-\bar{M}$. Surprisingly, the electronic band structure measurements suggest a structural relaxation occurring at 4.4° twist angle, and formation of large, untwisted bilayer regions.

# 1 Introduction

After the isolation of graphene [1], it became clear that several other materials can be stable in single layer form, and that the material's thickness plays a crucial role in determining the properties [2]. Since then, numerous two-dimensional (2D) materials have been isolated, showcasing significant potential for diverse technological applications [3], with transition metal dichalcogenides (TMDs) coming to the fore as a particularly interesting class.

A defining characteristic of 2D materials is that the covalent bonding within the plane is much stronger than the van der Waals interaction between planes, a feature allowing the isolation of single layers by mechanical exfoliation. The single layers are also easily stackable, enabling the realization of homo- and heterostructures where the interactions generate emergent properties not present in the precursor materials [4]. This makes stacking [5, 6] one of the ways to induce novel properties in 2D materials, next to gating [7], or straining [8].

When two layers are stacked, a small mismatch between the two lattices in the heterobilayers or a small twist angle in the homobilayers gives rise to a moiré pattern - a long-distance periodicity. This superlattice can significantly affect the electronic properties of the material and, under certain conditions, host a plethora of diverse and unique effects [9] because the different moiré periodic potentials modulate the electronic band structure due to correlation effects. Bilayer graphene with specific twist angles known as magic angles, exhibits an electronic band structure with flat bands where the electrons are extremely localized and, contingent on band filling [6], the whole phase diagram of electronic states can be explored [10], from a superconducting domain [5, 11], to Mott insulating states [6, 12], or ferromagnetic behavior [13]. These phenomena have been extensively studied in the last few years [9], and direct observation of the flat bands *via* angle-resolved photoemission spectroscopy (ARPES) has recently been achieved [14, 15]. In fact, due to the short mean free path of photoelectrons excited from the sample, ARPES is predominantly sensitive to the topmost layers of the material and this, combined with the ability to directly image the momentum-resolved band structure, makes it an ideal technique to study the electronic band structure of single or few-layer systems.

The above-mentioned findings for twisted bilayer graphene founded the field of twistronics and led to a better understanding of strongly correlated systems [5, 6]. In order to advance this field, however, it is imperative to analyze the behavior of other 2D materials such as TMDs, in which many-body phases have been recently unveiled, sparking a growing scientific interest [16].

Transition metal dichalcogenides are a class of materials composed of a transition metal plane (*e.g.* Mo or W) sandwiched between two chalcogen atom planes (*e.g.* S or Se). Single three-atom-thick TMD layers show exciting physical properties, such as a direct band gap and high flexibility [17, 18], not attainable in their bulk counterparts. Among TMDs, $WS_2$ has garnered specific attention holding great promise for next generation opto-electronic devices [19], and offering new avenues for band gap engineering through strain [20] and heterojunction formation [21]. Theoretical predictions for twisted bilayer TMDs have highlighted a high density of states in their flat bands [22, 23], along with the existence of topological excitons [24]. Recent studies foresee rich correlated effects close to the 4° twist angle [25], including flat bands for $WSe_2$ at 3.48° [26], identifying this class of materials as a valuable testing ground for studying novel condensed matter phenomena like many-body interactions or topological phases [25].

On the experimental side moiré excitons have been extensively investigated in TMD heterostructures [27-31], and flat bands have been observed by scanning tunnelling spectroscopy and second harmonic generation in twisted bilayer TMDs [32, 33]. Despite the strong interest, to date, only a few ARPES measurements of the electronic band structure have been performed on twisted TMDs. Notable observations include moiré bands in $WSe_2$ with a 5.1° twist angle [34], flat bands in $WSe_2$ with a 57.4° twist angle [35], as well as band hybridization effects [36] and mini bands in TMD heterostructures [37]. However, direct measurements of the band structure of bilayer TMDs with twist angles close to 4°, validating the predicted rich correlation effects, [25] are still lacking.

The small size of twisted TMD samples calls for a good spatial resolution, particularly crucial when analyzing the regions on the order of a few micrometers where the moiré pattern is homogeneous. Consequently, employing a nanometer-sized probe becomes essential. In this article, we present a nano-ARPES investigation of the electronic band structure of bulk, single layer (SL) and twisted bilayer (TWBL) $WS_2$. Specifically, a twisted bilayer $WS_2$ with a twist angle of 4.4° falling within the angle range of interest for exploring correlated phenomena [25] was prepared and the ARPES results compared with density functional theory (DFT) calculations.

## 2 Experimental and theoretical details

ARPES measurements require atomically flat and conductive substrates. Accordingly, we opted for epitaxial graphene on silicon carbide (Gr/SiC) as a substrate. Our preparation method involved a combination of sublimation growth in an Ar atmosphere [38] and use of a polymer as an additional carbon source [39]. Following a wet-chemical cleaning procedure, the polymer (AZ5214E, MicroChemicals GmbH) was deposited onto the SiC surface *via* spin-coating. Subsequently the samples were heated in a custom-built furnace [40] to 900 °C in a 1000 mbar Ar atmosphere and the graphene growth was carried out at 1000 mbar with zero Ar gas flow at three temperature steps of 1200 °C, 1400 °C and 1750 °C [41]. Single layer $WS_2$ was grown by salt-assisted chemical vapor deposition, following the standard procedure [42] with S, $WO_3$ and NaCl powders as precursors (all of 99 % purity, Sigma Aldrich). The reaction took place in a custom-built quartz tube surrounded by two different heaters - a CARBOLITE GERO programmable furnace and a WATLOW heating belt - to ensure two temperature regions. For the $WO_3$ precursor, the furnace temperature was kept at 800-820 °C, while the heating belt around the sulfur source maintained it at 200 °C during $WS_2$ growth. The deposition was performed with Ar as a carrier gas, and the precursors were deposited on a Si wafer covered by 300 nm of thermally grown $SiO_2$ (University Wafer).

The transfer of the single layer $WS_2$ from the $SiO_2$/Si substrate onto the Gr/SiC substrate, and the preparation of the twisted bilayer, was performed by a polymer-assisted technique [43]. Specifically, a polymer stamp made of polydimethylsiloxane (PDMS, Sigma Aldrich), covered by polycarbonate (PC, Sigma Aldrich), was prepared on a glass slide. The PDMS/PC stamp was attached to the first $WS_2$ sample on $SiO_2$/Si, preheated to 130 °C, and then peeled off and attached to the target substrate of graphene on silicon carbide. Following this step, the stage was heated to 180 °C to melt the PC layer, facilitating the detachment of the stamp from the sample. To prepare the twisted bilayer $WS_2$, a new polymer stamp was used to pick up another $WS_2$ flake, which was then attached on top of the $WS_2$ flake on Gr/SiC prepared in the previously described step. Finally, after detaching the second stamp, the sample was rinsed in isopropanol to remove any residual polymer traces. As ARPES is highly surface-sensitive, ensuring an atomically clean sample surface is paramount for obtaining good quality data. To achieve this, after the deposition of each layer, the sample was carefully scanned with an atomic force microscope (AFM) using the tip to remove any bubbles or impurities which may end up trapped between the layers and could dramatically affect the quality of the spectroscopic measurements. The clean transfer procedure, together with an *in situ* 30 min annealing step at 150 °C prior to the ARPES measurement to desorb any water and hydrocarbons, warranted sample cleanliness.

Nano-ARPES measurements were conducted at the ANTARES beamline of the SOLEIL synchrotron, equipped with a hemispherical analyzer (MB Scientific, type A-1) on a spot with a diameter of 600 nm. All the measurements were carried out in ultra-high vacuum at a temperature of 70 K, using a photon energy of 100 eV and linearly horizontally polarized light. The energy resolution was 70 meV [44].

DFT calculations were performed using the Quantum Espresso simulation package [45, 46]. The electron-ion interactions were described *via* the projector-augmented wave method [47, 48], and the wave functions were expanded in a plane wave basis set using the cutoff of 800 eV. The Perdew-Burke-Ernzerhof (PBE) parametrization of the generalized gradient approximation was used as the exchange-correlation functional. A vacuum region of 20 Å along the c-axis was employed to separate the spurious images of mono- and bilayers, minimizing interactions between the system replicas. Lattice constants a= 3.18 Å and c= 12.98 Å (only relevant for the bulk) were used. The internal degrees of freedom were relaxed until the forces were below $10^{-3}$ Ry Bohr$^{-1}$, incorporating van der Waals corrections (DFT-D3) [49]. Brillouin zones (BZs) were sampled on k-meshes of 18×18×6 and 18×18×1 for the bulk and 2D layers, respectively. Spin-orbit coupling was taken into account self-consistently in all the simulations except for relaxations. Post-processing calculations of band structures and spin textures on dense k-meshes were performed using the PAOFLOW code [50, 51].

Atomic Force Microscopy is a scanning probe microscopy technique capable of resolving features at the nanometer scale. The specific AFM instrument employed in our study was a conductive-mode system manufactured by Bruker. This module is based on contact mode scanning and provides valuable topographical information as well as a map of the tunneling current passing through the sample's surface. During the scanning process, the conductive AFM (c-AFM) simultaneously records height, current, and friction signals. Samples on Gr/SiC were inspected after the dry transfer procedures described above. To obtain the moiré pattern from both the current and friction channels, we employed a conductive probe holder and tip.

## 3 Results and Discussion

To determine the twist angle between the two layers of WS$_2$ we performed conductive atomic force microscopy measurements [52]. The c-AFM image in Figure 1 shows two distinct regions of the sample: a single layer (SL) region at the top, and a twisted bilayer (TWBL) region at the bottom. The pattern visible in the twisted bilayer WS$_2$ is the moiré superstructure resulting from the interference between electron densities of the two individual layers. The wavelength of the moiré is defined as the distance between the two local current maxima (bright spots) and can be expressed as $\lambda_M = \frac{a}{2\sin(\vartheta/2)}$, where $a$ is the lattice constant of WS$_2$ ($a = 3.18$ Å) and $\vartheta$ the twist angle between the layers. In our sample, the moiré length was found to be $\lambda_M = 4.1 \pm 0.1\ nm$, corresponding to a twist angle of $\vartheta = 4.4 \pm 0.2°$ between the two WS$_2$ layers. Further information about the periodicity of the moiré pattern can be found in the Supplementary Information Figure S1.

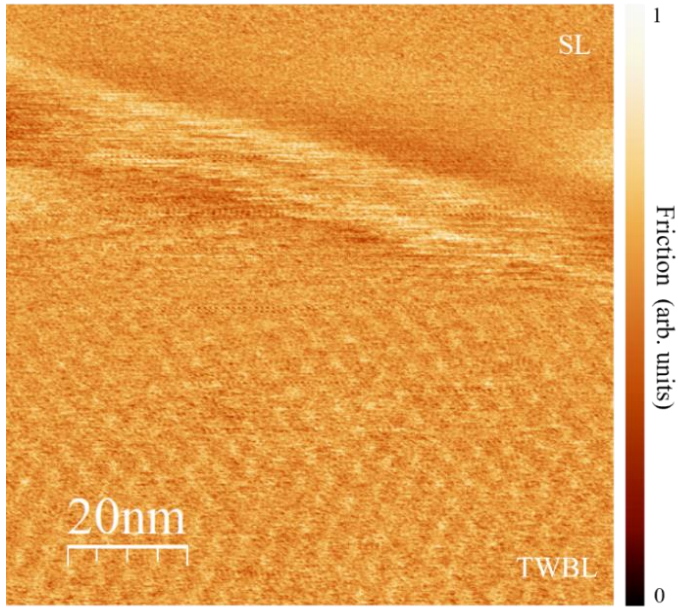

**Fig. 1** Conductive atomic force microscopy (c-AFM) image showing a single layer (SL) WS$_2$ region at the top, and a twisted bilayer (TWBL) WS$_2$ region at the bottom. The bright line is the edge between the SL and the TWBL, while the pattern visible in the TWBL region is the moiré superstructure given by the interference of electron densities.

Subsequently ARPES measurements were performed on this same sample, as well as on bulk WS$_2$, to comprehensively characterize the electronic structure, and discern changes arising from the twist between the layers. The data for bulk WS$_2$ are reported in Figure 2. The hexagonal surface symmetry requires the investigation of the band dispersion along two high-symmetry directions, $\bar{\Gamma} - \bar{K}$ and $\bar{\Gamma} - \bar{M}$. As can be seen in Figure 2, the valence band maximum (VBM) is positioned around the $\bar{\Gamma}$ point, consistent with expectations [53]. Multiple branches are discernable, as typical for a bulk semiconducting TMD [54]. Notably, we observe good agreement between the experimental data and our DFT calculations, represented by the red overlay. This alignment confirms the reliability of our theoretical model in capturing the electronic properties of bulk WS$_2$. We proceeded with the WS$_2$ sample on Gr/SiC, where we collected ARPES data both from the SL region, which we present in Figure 3, and from the TWBL region, which we show in Figure 4. The precise ARPES measurement location marked on the spatially resolved ARPES map can be found in the Supplementary Information, Figure S2.

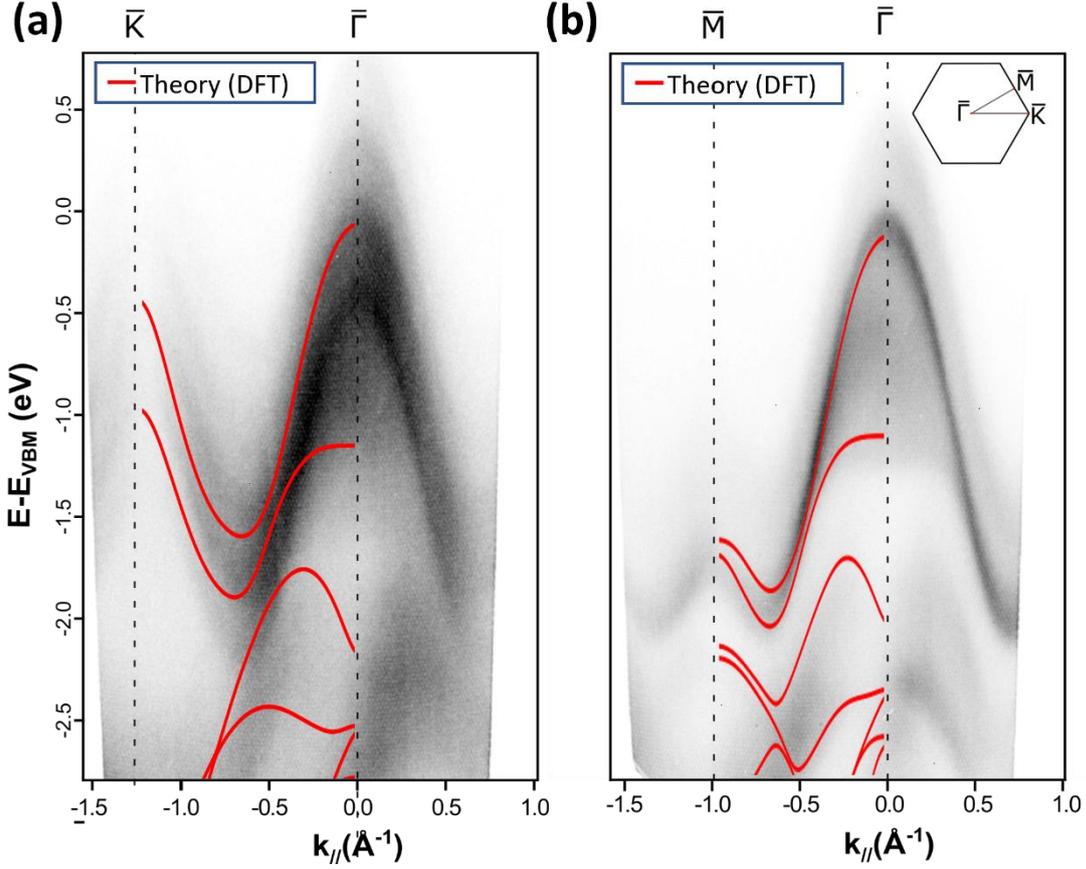

**Fig. 2** Energy-momentum spectra presenting the band structure of bulk WS$_2$ along the two high-symmetry directions $\bar{\Gamma} - \bar{K}$ (a) and $\bar{\Gamma} - \bar{M}$ (b). Corresponding DFT calculations are shown in red. The high-symmetry directions of interest are marked in the Brillouin zone sketched in the inset, (b) top right.

Figure 3 illustrates the SL WS$_2$ band structure acquired along the high-symmetry directions $\bar{\Gamma} - \bar{K}$ and $\bar{\Gamma} - \bar{M}$. Our measurements confirm the SL nature, evidenced by the presence of only one band at the $\bar{\Gamma}$ point, in contrast to the bulk data where numerous bands merge into a continuum. Consistent with expectations for SL WS$_2$, we identify the VBM at the $\bar{K}$ point [55]. A careful look at the VBM reveals that the intensity is very low at the $\bar{K}$ point, attributable to unfavorable matrix element effects in our specific measurement geometry. This matrix element effect is particularly pronounced along the $\bar{\Gamma} - \bar{K}$ high-symmetry direction, while it is less noticeable along the $\bar{\Gamma} - \bar{M}$ high-symmetry direction. Furthermore, our data show electronic bands coming from the WS$_2$ single layer alongside graphene bands, denoted by the gray arrows in Figure 3. No hybridization between the graphene and WS$_2$ is observed, indicating a weak interaction between WS$_2$ and the Gr/SiC substrate.

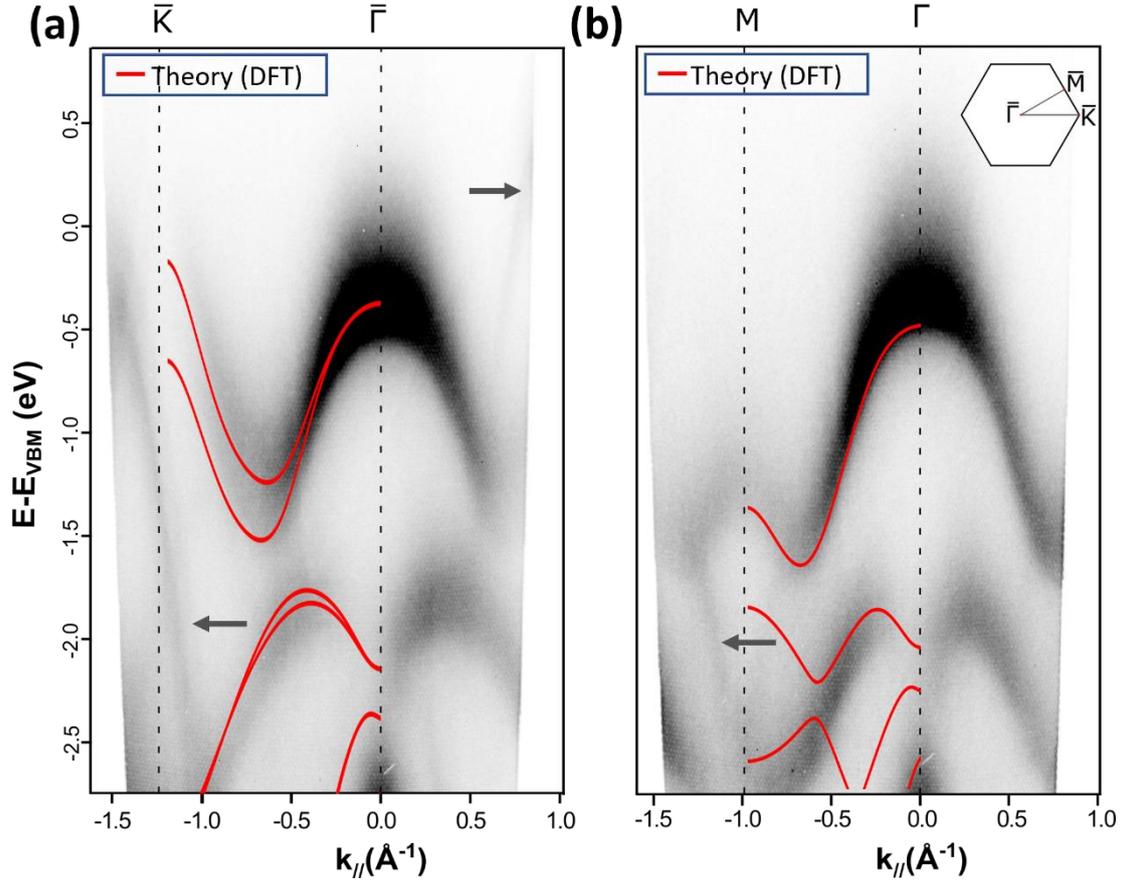

**Fig. 3** Energy-momentum spectra presenting the band structure of single layer WS$_2$ along the two high-symmetry directions $\overline{\Gamma} - \overline{K}$ (a) and $\overline{\Gamma} - \overline{M}$ (b). The corresponding DFT calculations are shown in red. The high-symmetry directions of interest are marked in the Brillouin zone sketched in the inset, (b) top right. For single layer WS$_2$, the valence band maximum is located at the $\overline{K}$ point. Gray arrows indicate the graphene bands that are visible because the WS$_2$ layer is placed on a graphene/SiC substrate.

Comparison of the experimental band structure of the single layer WS$_2$, measured by nano-ARPES, with the theoretically calculated bands using DFT demonstrates excellent agreement. This consistency extends to the states at the $\overline{\Gamma}$ and the $\overline{K}$ points, as well as the bands lying at higher binding energy. However, an apparent discrepancy arises for the lower-lying bands along the $\overline{\Gamma} - \overline{M}$ direction, Figure 3(b), where DFT predicts a gap due to spin-orbit coupling that is not evident in the experimental data. Careful examination of this region reveals the presence of a band gap, with further details provided in Supplementary Information Figure S3.

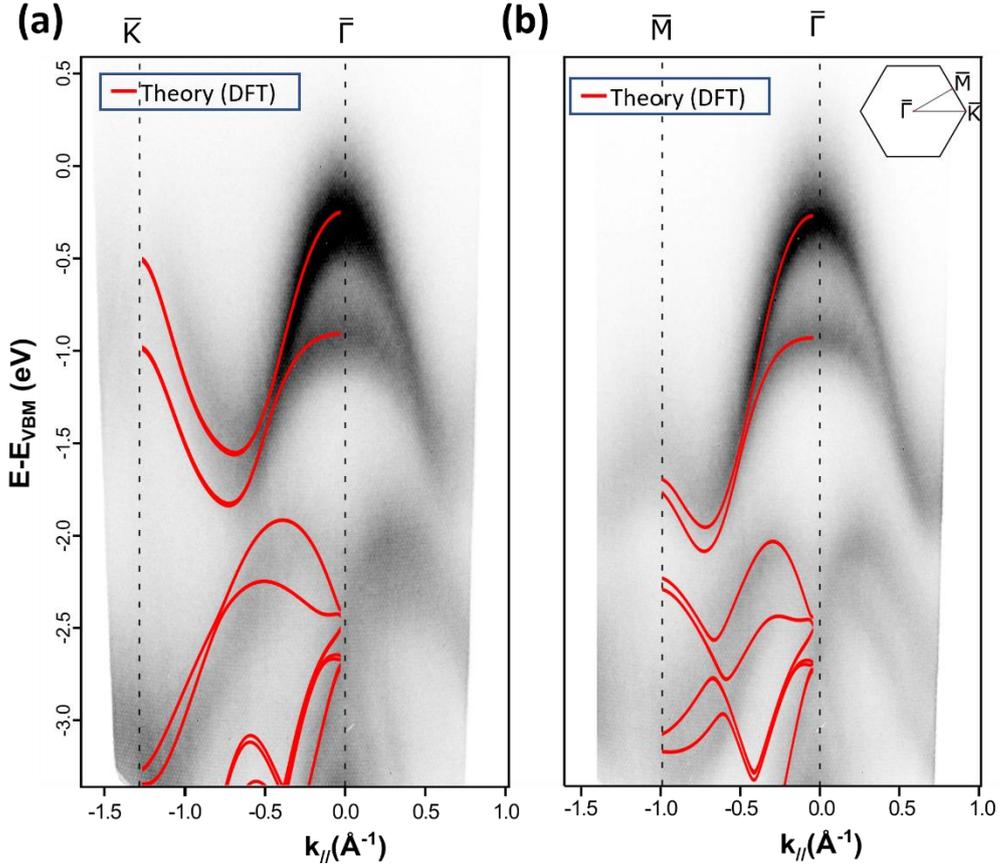

**Fig. 4** Energy-momentum spectra presenting the band structure of bilayer WS$_2$ twisted at a twist angle of 4.4° along the two high-symmetry directions $\bar{\Gamma} - \bar{K}$ (a) and $\bar{\Gamma} - \bar{M}$ (b). DFT calculations for AA′ stacking, one of possible WS$_2$ bilayer stackings without a twist angle, are shown in red. The high-symmetry directions of interest are marked in the Brillouin zone sketched in the inset, (b) top right.

In Figure 4, the electronic band structure of TWBL WS$_2$ at 4.4° twist angle is depicted along the high-symmetry directions $\bar{\Gamma} - \bar{K}$ and $\bar{\Gamma} - \bar{M}$, mirroring the measurements performed for the single layer. Notably, despite theoretical predictions for TMDs with a twist angle close to 4° [25] indicating the presence of a mini Brillouin zone (BZ) and flat bands at the $\bar{\Gamma}$ point, our experimental data do not reveal such signatures in the electronic band structure. To further scrutinize the possibility of flat bands, we conducted a 2D curvature investigation, as demonstrated in Ref [56]; the results are shown in Supplementary Information Figure S4. However, even in the 2D curvature enhanced data the absence of flat bands at the $\bar{\Gamma}$ point persists, raising intriguing questions about the nature of the observed electronic structure.

To delve into these questions, we turned to DFT calculations, exploring various bilayer WS$_2$ configurations. The calculations that best agree with our experimental band structure probed by ARPES are presented in Figure 4. Remarkably, our experimental data aligns excellently with the theoretical calculations conducted for a regularly stacked (untwisted) bilayer WS$_2$, specifically in the *AA′* stacking configuration, where the W atoms align over the S atoms, and the S atoms are fully eclipsed [57] (see Supplementary Information Figure S5 for a sketch of this stacking). Similar agreement is observed for the *AA* and *AB* stacking configurations, as shown in Supplementary Information Figures S6 and S7.

This outcome parallels recent findings of photoluminescence measurements on a CVD-grown MoSe$_2$/WSe$_2$ heterostructure, where the coexistence of both moiré and moiré-free regions was explored [58]. Analogously, an investigation by scanning electron microscopy of mesoscopic reconstruction in MoSe$_2$/WSe$_2$ heterobilayers at small twist angles up to 3° revealed the presence of reconstructed domains below 100 nm in size [59]. These studies underline the impact of finite elasticity, where lattice deformation may occur due to the competition between intralayer strain and interlayer adhesion energy, resulting in the expansion of the most energetically favored structure and the formation of hexagonal or triangular domains. AFM investigations of homobilayers with a twist below 2° further support this picture [60].

The concept of relaxation domains finds theoretical support in recent work exploring lattice relaxation in twisted bilayer TMDs at small twist angles [61]. Arnold *et al.* [61] demonstrated how the properties of TWBL MoS$_2$, akin to our WS$_2$, undergo a continuous transition with varying twist angle between the layers and no single magic angle exists, where the transition between structural regimes occurs. In the soliton regime (0° to 3° for the domain-soliton regime and 3° to 6° for the soliton regime in MoS$_2$), lattice reconstruction occurs due to the corrugation of individual layers, leading to the formation of large domains predominantly consisting of relaxed (untwisted) bilayers. Our twist angle of 4.4° fits perfectly within this window, suggesting that large domains of relaxed WS$_2$ bilayers exhibiting regular stacking contribute significantly to our ARPES signal due to the averaging nature of ARPES coupled with the substantial size of these domains.

It is crucial to acknowledge that, despite employing nano-ARPES, the investigated region encompasses the entire area of the beam spot, which in our case is 600 nm in diameter. This area significantly exceeds the dimensions of individual domains, which, as observed in the case of MoS$_2$ [61], are on the order of a few nm [61]. Consequently, the resulting band structure measurement is a weighted average of multiple domains within the beam spot. Notably, the domains where the structure is relaxed are considerably larger than the smaller boundary regions (solitons). Given this, the majority of signal in our ARPES measurements originates from the relaxed domains, and the band structure we observe resembles that of a bilayer without a twist. In this context of atomic reconstruction [61], it becomes apparent why our ARPES measurement exhibits no effect of the correlated states, and, instead, demonstrates excellent agreement with DFT calculations for the regularly stacked (untwisted) bilayer.

## 4 Conclusions

In this study we conducted a comprehensive nano-ARPES investigation of WS$_2$ to elucidate the nature of the electronic structure of WS$_2$ bilayers with a small twist angle. Measurements encompassed bulk, single layer and twisted bilayer WS$_2$ with a twist angle of 4.4°, for which nano-ARPES spectra were acquired along the main high-symmetry directions $\bar{\Gamma} - \bar{K}$ and $\bar{\Gamma} - \bar{M}$.
Contrary to earlier theoretical predictions anticipating correlated states for a 4.4° twist angle, our results reveal a surprising behavior. The electronic structure closely resembles that of an untwisted, regular WS$_2$ bilayer, aligning with recent theoretical propositions, which suggest that for specific twist angle intervals, including 4.4°, atomic relaxation occurs, leading to the formation of large domains of lower energy, *i.e.* untwisted bilayer regions. Such an interpretation is further supported by our DFT calculations, which showcase a remarkable agreement between the calculations for regularly stacked WS$_2$ and our experimental data. This is the first time that this effect has been observed in ARPES measurements, opening the door to further exploration of relaxation effects and their signatures in the electronic structure. Our finding underlines the significance of comprehending the fundamental process involved in these reconstruction effects, especially as reconstruction can be undesirable when aiming for a twisted system with specific properties and highlights ARPES as a tool well suited to provide insights into this.

## Acknowledgments

J.S. acknowledges a Rosalind Franklin Fellowship from the University of Groningen and funding from NWO under the contract OCENW.M.22.063. Part of the work was supported by the National Science Centre, Poland, under Grant No. 2018/30/E/ST5/00667. The calculations were carried out on the Dutch national e-infrastructure with the support of the SURF Cooperative (EINF-5312) and on the Hábrók high-performance computing cluster of the University of Groningen. We acknowledge SOLEIL for the provision of synchrotron radiation facilities, where nano-ARPES measurements were performed on the ANTARES beamline under proposal number 20201396.

# Supporting Information for
# Nano-ARPES investigation of twisted bilayer tungsten disulfide

Giovanna Feraco[1], Oreste De Luca[1†], Przemysław Przybysz[1,2], Homayoun Jafari[1], Oleksandr Zheliuk[1], Ying Wang[1], Philip Schädlich[3], Pavel Dudin[4], José Avila[4], Jianting Ye[1], Thomas Seyller[3], Paweł Dąbrowski[2], Paweł Kowalczyk[2], Jagoda Sławińska[1*], Petra Rudolf[1*], Antonija Grubišić-Čabo[1*].

[1]Zernike Institute for Advanced Materials, University of Groningen, Nijenborgh 4, 9747AG Groningen, The Netherlands

[2]Faculty of Physics and Applied Informatics, University of Łódź, Pomorska 149/153, 90-236 Łódź, Poland

[3]Institut für Physik, Technische Universität Chemnitz, Reichenhainer Str. 70, 09126 Chemnitz, Germany

[4]Synchrotron SOLEIL, Université Paris-Saclay, Saint Aubin, BP 48, 91192 Gif sur Yvette, France

*Corresponding authors. E-mail: a.grubisic-cabo@rug.nl, p.rudolf@rug.nl, jagoda.slawinska@rug.nl. AGC and PR for ARPES measurements; JS for DFT calculations.

## Twist angle determination with AFM

Atomic force microscopy (AFM) was utilized to determine the twist angle of the bilayer $WS_2$, as depicted in Figure S1 (a). A 2D Fast Fourier Transform (2D-FFT) was performed on the AFM image displayed in Figure S1 (a), and the results are shown in Figure S1 (b). The average length along all three axes, 4.1nm, was then taken as the reciprocal lattice length, which was used to calculate the wavelength of the moiré pattern in real space.

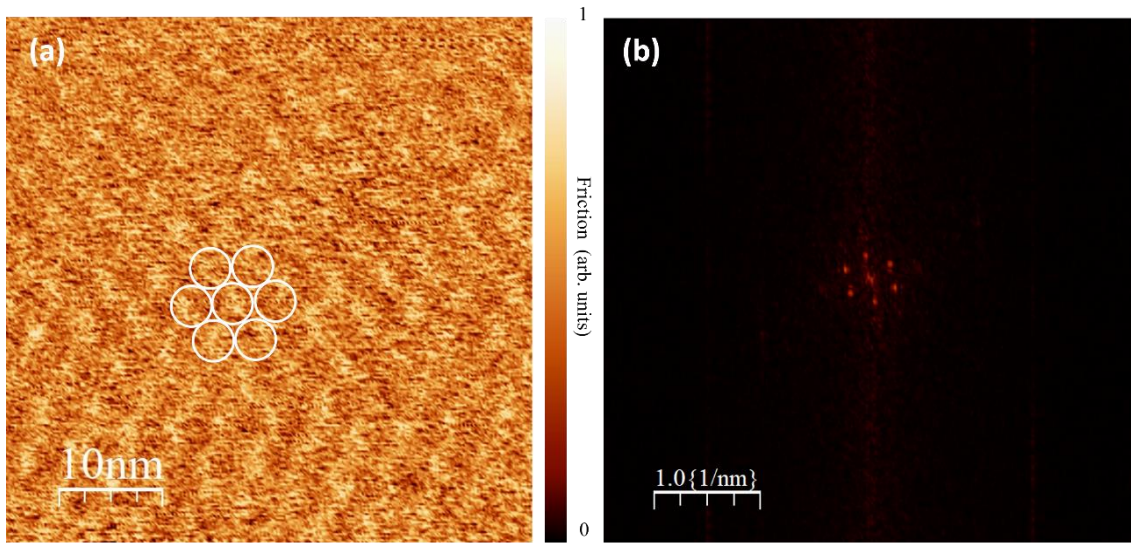

**Figure S1** (a) Conductive AFM (c-AFM) image displaying the characteristic moiré pattern in the twisted bilayer region; (b) 2D Fast Fourier Transform of the moiré pattern, highlighting hexagonal symmetry.

**ARPES mapping:** Prior to conducting angle-resolved photoemission spectroscopy (ARPES) measurements, an ARPES map of the flake (Figure S2) was created to identify the specific positions for the measurements; the selected positions are marked in Figure S2.

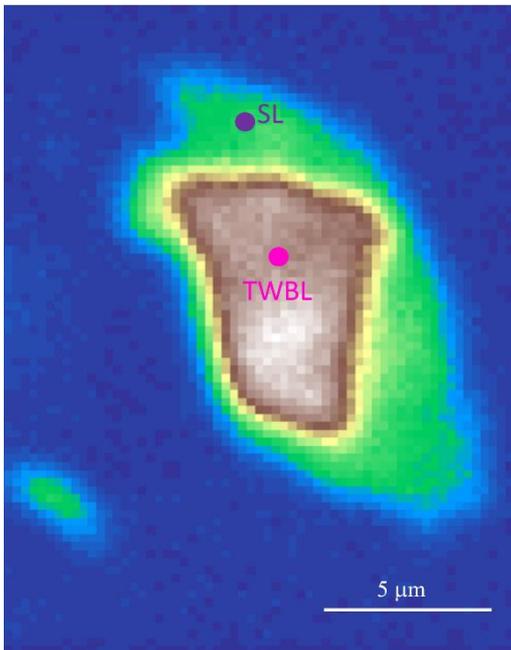

**Figure S2** Spatially resolved ARPES map showing different contrast for regions where single layer (SL) WS$_2$ and twisted bilayer (TWBL) WS$_2$ are located. The markers show the region where ARPES measurements were acquired for SL (purple) and TWBL (pink).

**Additional ARPES analysis:** Additional ARPES analysis was carried out to gain a deeper understanding of the electronic band structure of bilayer WS$_2$. This analysis revealed a previously unresolved band gap in the region where an apparent discrepancy between experimental results and theoretical calculations exists, Figure S3 (also see Figure 3 (b) in the main text).

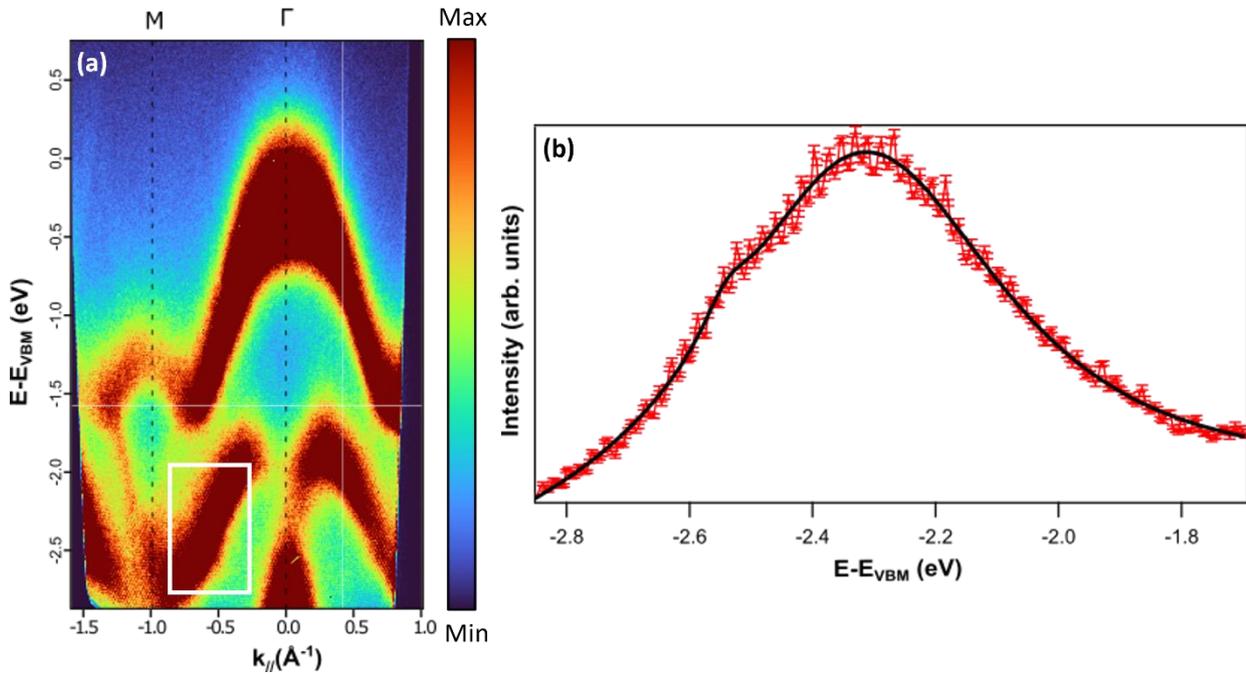

**Figure S3** (a) Energy-momentum spectrum of the single layer WS$_2$ along the $\overline{\Gamma} - \overline{M}$ high-symmetry direction with high contrast color scale to better highlight the presence of two different bands around -2.5 eV. The same data is shown in the main text, Figure 3(b) in linear grayscale; (b) Energy distribution curve (EDC) of WS$_2$ taken in the region around -2.5 eV, where a band gap in the deeper lying states is expected to show two contributions at different energies. This EDC confirms the presence of a bandgap that could not be resolved because of the limited energy resolution.

## 2D curvature of twisted bilayer WS$_2$

A 2D curvature method [1] was applied to the ARPES data of twisted bilayer WS$_2$ along the $\overline{K} - \overline{\Gamma} - \overline{K}$ direction in order to enhance weak features, such as possible mini-gaps and flat bands arising from the moiré superpotential. No flat bands are observed in the 2D curvature data, Figure S4, supporting the picture of lattice relaxation taking place in our system.

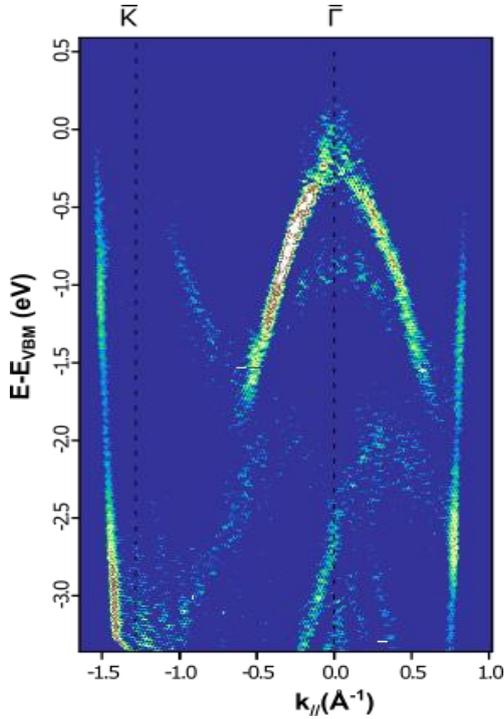

**Figure S4** Energy-momentum spectra showing the 2D curvature investigation of the band structure of twisted bilayer WS$_2$ at a twist angle of 4.4° along the high-symmetry direction $\overline{\boldsymbol{\Gamma}} - \overline{\boldsymbol{K}}$. While bands coming from the top and the bottom WS$_2$ layer are clearly resolved, we do not observe minibands arising from the moiré superpotential.

## Comparison of ARPES data with DFT calculations

Density functional theory (DFT) calculations for three possible WS$_2$ bilayer stackings were compared with the ARPES data in Figures S5, S6, and S7. These stackings include AA' with W atoms on top of S atoms (Figure S5), AA with W atoms on top of W atoms and S atoms over S atoms (Figure S6), and the staggered AB stacking with S atoms on top of W atoms (Figure S7). The data shows comparable agreement for all three stackings.

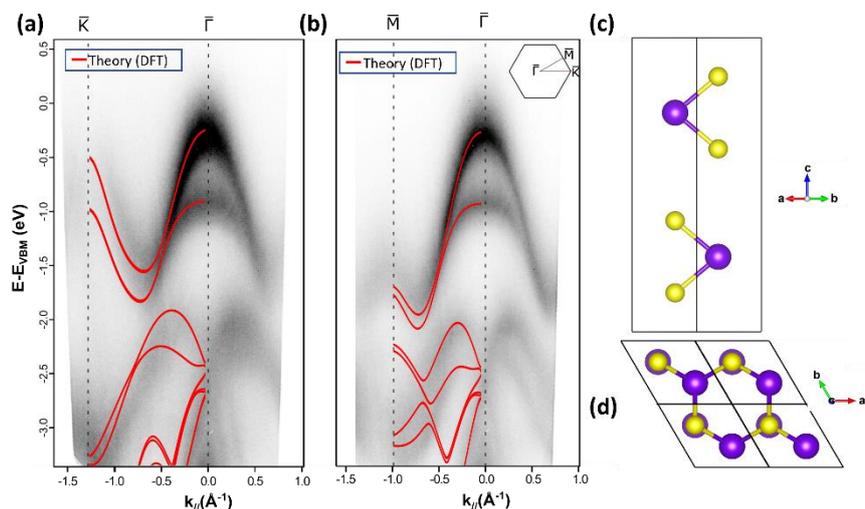

**Figure S5** Comparison of the experimental ARPES data with the electronic structure oh AA' stacked WS$_2$. Energy-momentum spectra showing the band structure of twisted bilayer WS$_2$ at a twist angle of 4.4° along the two high-symmetry directions $\overline{\Gamma} - \overline{K}$ (a) and the $\overline{\Gamma} - \overline{M}$ (b). The DFT calculation shown in red is for the AA' stacking. The high-symmetry directions used for the measurements are marked in the Brillouin zone sketched in the inset (b) top right; a side view (c) and a top view (d) of the AA' stacking.

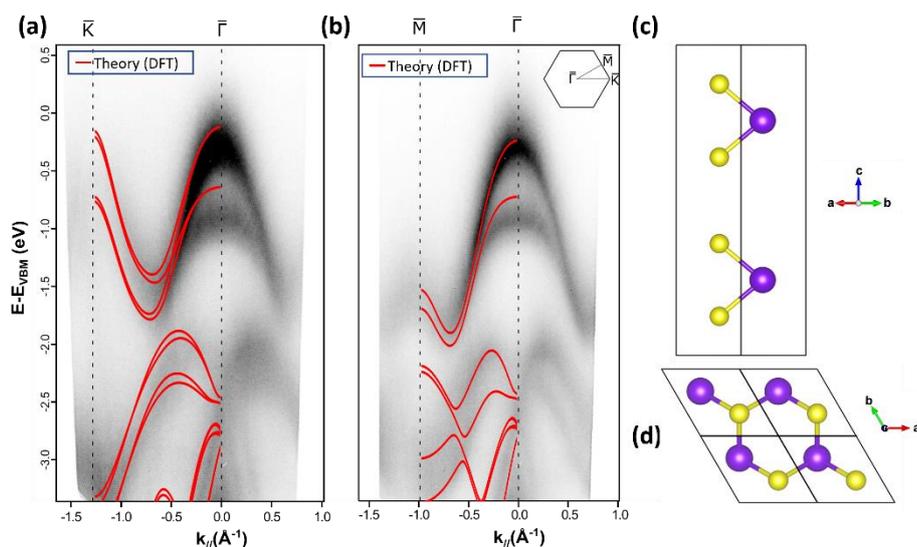

**Figure S6** Comparison of the experimental ARPES data with the electronic structure of AA stacked WS$_2$. Energy-momentum spectra showing the band structure of twisted bilayer WS$_2$ at a twist angle of 4.4° along the two high-symmetry directions $\overline{\Gamma} - \overline{K}$ (a) and the $\overline{\Gamma} - \overline{M}$ (b). The DFT calculation is shown in red is for the AA stacking. The high-symmetry directions used for the measurements are marked in the Brillouin zone sketched in the inset (b) top right; a side view (c) and a top view (d) of the WS$_2$ AA stacking.

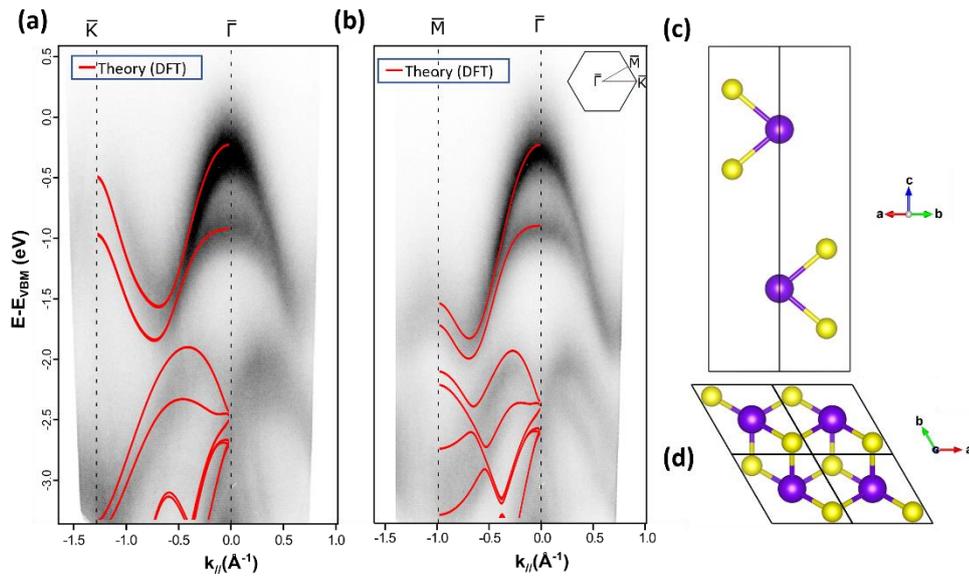

**Figure S7** Comparison of the experimental ARPES data with the electronic structure of AB stacked WS$_2$. Energy-momentum spectra showing the band structure of twisted bilayer WS$_2$ at a twist angle of 4.4° along the two high-symmetry directions $\overline{\Gamma} - \overline{K}$ (a) and the $\overline{\Gamma} - \overline{M}$ (b). The DFT calculation shown in red is for the AB stacking. The high-symmetry directions used for the measurements are marked in the Brillouin zone sketched in the inset (b) top right; a side view (c) and a top view (d) of the AB stacking.